# Restoration of Lorentz Invariance of 't Hooft-Polyakov Monopole Field

K. Rasem Qandalji

*Amer Institute*
*P.O. Box 1386, Sweileh, 11910*
*JORDAN*
*E-mail:* qandalji@hotmail.com

**ABSTRACT**

Lorentz invariance is broken for the non-Abelian monopoles. Here we will consider the case of 't Hooft-Polyakov monopole and show that the Lorentz invariance of its field will be restored using Dirac quantization.

## 1. Introduction

Soon after the non-Abelian monopoles were shown to break color [1],[2],[3], Balachandran *et al* [4] showed that monopoles also break Lorentz invariance. They showed that to be true for topologically stable as well as unstable monopoles: In the former case, the monopoles are predicted as stable topological excitations by gauge theories based on a simply connected gauged group *G*, which is broken spontaneously, by the "Higgs vacuum" (defined by Eqs. (2.1,2) below), to a subgroup *H* which is not simply connected. *H* cannot be simply connected since classes of its first homotopy group, $\Pi_1(H)$, are isomorphic to the topological quantum numbers of the magnetic charge. If $\Pi_1(H) = 0$ then there can be no magnetic monopole: For *G* simply connected we have, $\Pi_1(H) \simeq \Pi_2(G/H)$ where the right coset *G/H* is isomorphic to the vacuum manifold of the Higgs field $\mathcal{M}_o$ [5]. Balachandran and collaborators also showed that the Lorentz invariance is broken in the case of topologically unstable magnetic monopoles arising from the GNO configurations (GNO configurations are named after Goddard, Nuyts, and Olive who first introduced them [6].)

In this article we will consider the 't Hooft-Polyakov monopole's field [7],[8] (outside its core, i.e. in the Higgs vacuum region) and show that using results from the Dirac quantization of this field [9] will help restoring the Lorentz invariance broken at the classical level.

## 2. Preliminaries

**'t Hooft-Polyakov monopole [5] and the Dirac Quantization of its field [9].**

(We will use the metric (+,-,-,-). Index with Greek alphabet runs from 0 to 3, and a Latin alphabet index runs from 1 to 3, unless otherwise stated.)

The 't Hooft-Polyakov monopole model consists of an *SO*(3) gauge field interacting with an isovector Higgs field $\phi$. The model's Lagrangian is:

$$\mathcal{L} = -\frac{1}{4} G_a^{\mu\nu} G_{a\mu\nu} + \frac{1}{2} D^\mu \phi . D_\mu \phi - V(\phi),$$

where $\phi = (\phi_1, \phi_2, \phi_3)$, and $V(\phi) = \frac{1}{4}\lambda(\phi_1^2 + \phi_2^2 + \phi_3^2 - a^2)$. $G_a^{\mu\nu}$, is the gauge field strength: $G_a^{\mu\nu} = \partial^\mu W_a^\nu - \partial^\nu W_a^\mu - e\varepsilon_{abc} W_b^\mu W_c^\nu$, where $W_a^\mu$ is the gauge potential.

The model's Lagrangian full symmetry Group *SO*(3), generated by $T_a$'s, is spontaneously broken, by the Higgs Vacuum (defined below), down to *SO*(2) ($\simeq U(1)$), generated by $\frac{\phi.\mathbf{T}}{a}$. The model's non-singular extended solution looks, at large distances, like a Dirac monopole.

The monopole's energy finiteness implies that there is some radius $r_0$ such that for $r \geq r_0$ we have, to a good approximation:

$$D^\mu \phi \equiv \partial^\mu \phi - e\mathbf{W}^\mu \times \phi = 0, \qquad (2.1)$$

and $\qquad \phi_1^2 + \phi_2^2 + \phi_3^2 - a^2 = 0, \; (\Rightarrow V(\phi) = 0). \qquad (2.2)$

Regions of space-time, where the above two equations are satisfied, constitute the Higgs Vacuum.

The general form of $\mathbf{W}^\mu$ in the Higgs Vacuum is [10]:

$$\mathbf{W}^\mu = \frac{1}{a^2 e} \phi \times \partial^\mu \phi + \frac{1}{a} \phi A^\mu, \qquad (2.3)$$

where $A^\mu$ is arbitrary.

It follows that:

$$\mathbf{G}^{\mu\nu} = \frac{1}{a} \phi F^{\mu\nu} \qquad (2.4)$$

where, $\qquad F^{\mu\nu} = \frac{1}{a^3 e} \phi . (\partial^\mu \phi \times \partial^\nu \phi) + \partial^\mu A^\nu - \partial^\nu A^\mu \qquad (2.5)$

So in Higgs vacuum: $\mathcal{L} = -\frac{1}{4} G_a^{\mu\nu} G_{a\mu\nu}$, and on account of (2.2,4) we get: $\mathcal{L} = -\frac{1}{4} F^{\mu\nu} F_{\mu\nu}$.

In the Higgs vacuum region, we also have the conjugate momentum of dynamical coordinates, $A^{\eta}(\mathbf{x})$'s and $\phi_i(\mathbf{x})$'s, given by [9]:

$$\Pi_{\eta}(x) \equiv \frac{\partial \mathcal{L}}{\partial \dot{A}^{\eta}(x)} = \frac{\varepsilon_{rst}}{a^3 e} \phi_r \partial_{\eta} \phi_s \partial_0 \phi_t + \partial_{\eta} A_0 - \partial_0 A_{\eta} = \begin{cases} 0 & \text{, for } \eta = 0 \\ F_{i0} & \text{, for } \eta = i = 1,2,3 \end{cases},$$

(2.6)

and $\quad \pi_l(x) \equiv \dfrac{\partial \mathcal{L}}{\partial \dot{\phi}_l(x)} = \dfrac{\varepsilon_{ijl}}{a^3 e} \phi_i \partial^k \phi_j \left( \dfrac{\varepsilon_{rst}}{a^3 e} \phi_r \partial_0 \phi_s \partial_k \phi_t + \partial_0 A_k - \partial_k A_0 \right).$ (2.7)

As for the *Dirac quantization of the monopole's field* (i.e. in the Higgs vacuum region), the details are given in ref.[9], but we will quote here the equations we will need in section **3.** below.

The complete set of constraints in the axial gauge, $\zeta_\alpha$, $(\alpha = 1,...,8)$ are [9]:

$$\zeta_1 = \phi_2 \Phi_1 - \phi_1 \Phi_2 - \frac{\alpha_3}{2} \chi \approx 0,$$

$$\zeta_2 = \phi_3 \Phi_2 - \phi_2 \Phi_3 - \frac{\alpha_1}{2} \chi \approx 0,$$

$$\zeta_3 = \frac{1}{2a^2}(\phi_1 \Phi_1 + \phi_2 \Phi_2 + \phi_3 \Phi_3) \approx 0,$$

$$\zeta_4 = \chi \equiv \phi_1^2 + \phi_2^2 + \phi_3^2 - a^2 \approx 0,$$

$$\zeta_5 = \partial^i \Pi_i \approx 0,$$

$$\zeta_6 = \frac{1}{ae}(\phi_2 \partial^3 \phi_1 - \phi_1 \partial^3 \phi_2) - A^3 \phi_3 \approx 0,$$

$$\zeta_7 = \frac{1}{ae}(\phi_3 \partial^3 \phi_2 - \phi_2 \partial^3 \phi_3) - A^3 \phi_1 \approx 0,$$

$$\zeta_8 = A^3 \approx 0,$$

(2.8)

where $\Phi_l \equiv \pi_l + \dfrac{\varepsilon_{ijl}}{a^3 e} \phi_i \partial^k \phi_j \Pi_k$, and $\alpha_k \equiv \dfrac{3}{a^3 e} \Pi_j \partial^j \phi_k$.

It is also sufficient for our purposes here to mention that, the only non-vanishing elements of $C_{\alpha\alpha'}^{-1}$ are: $C_{16}^{-1}, C_{17}^{-1}, C_{18}^{-1}, C_{26}^{-1}, C_{27}^{-1}, C_{28}^{-1}, C_{34}^{-1}, C_{56}^{-1}, C_{57}^{-1}, C_{58}^{-1}$ and their transposes. Again, for the exact values of $C_{\alpha\alpha'}^{-1}$ in the Higgs vacuum region of the monopole see ref. [9].

# 3. Restoration of Lorentz Invariance

In this section we will show that incorporating quantum effects into the theory through evaluating the Dirac brackets [11],[12] of the Lorentz generators, using results quoted in sec.2, will result in the manifest restoration of Lorentz invariance of the monopole's field which was broken at the classical level.

The conventional expressions of the angular momenta and boosts for the Yang-Mills fields are,

$$L_i = \int d^3x \left[ \mathbf{x} \times (\mathbf{E}_a \times \mathbf{B}_a) \right]_i,$$
$$K_i = \frac{1}{2} \int d^3x \, x_i \left( E_{ja} E_{ja} + B_{ja} B_{ja} \right), \tag{3.1}$$

where $a$ is the internal symmetry index.

We also have:

$$(\mathbf{B}_a)_i \equiv \frac{1}{2} \varepsilon_{ijk} G_{ajk},$$
$$(\mathbf{E}_a)_i \equiv -G_{a0i}. \tag{3.2}$$

In the monopole's field outside its core (i.e. in the Higgs vacuum region) and by using Eqs. (2.2), (2.4), (2.5), (2.6), (3.1), and (3.2), $L_i$ and $K_i$ will reduce there to

$$L_i = -\frac{1}{2} \varepsilon_{ijk} \varepsilon_{klm} \varepsilon_{mpq} \int d^3x \, x_j F_{0l} F_{pq}$$
$$K_i = \frac{1}{2} \int d^3x \, x_i \left( F_{0j} F_{0j} + \frac{1}{4} \varepsilon_{lpq} \varepsilon_{lkm} F_{pq} F_{km} \right) \tag{3.3}$$

**a.** First, we evaluate the (equal time) Dirac bracket of two $L_i$'s [11],[12]:

$$\{L_i(t), L_h(t)\}_{D(\zeta)} \equiv \{L_i(t), L_h(t)\} - \int \int \{L_i(t), \zeta_\alpha(\mathbf{x},t)\} \, d^3x \, C^{-1}_{\alpha\alpha'}(\mathbf{x}, \mathbf{x}';t) \, d^3x' \{\zeta_{\alpha'}(\mathbf{x}',t), L_h(t)\} \tag{3.4}$$

Using Eq. (2.6) the (equal time) first term on the right hand side of Eq. (3.4) will be,

$$\{L_i(t), L_h(t)\} = \frac{1}{4} \varepsilon_{ijk} \varepsilon_{klm} \varepsilon_{mpq} \varepsilon_{hgf} \varepsilon_{fed} \varepsilon_{dcb} \int \int d^3x \, d^3x' \, x_j x'_g \left\{ \Pi_l(x) F_{pq}(x), \Pi_e(x') F_{cb}(x') \right\}\Big|_{t'=t} \tag{3.5}$$

Using Eq. (2.5) we form :

$$\left\{ \Pi_l(x) F_{pq}(x), \Pi_e(x') F_{cb}(x') \right\}\Big|_{t'=t} = \Pi_e(x')\Big|_{t'=t} F_{pq}(x) \left( g_{cl} \partial_{b'} \delta(\mathbf{x}' - \mathbf{x}) - g_{lb} \partial_{c'} \delta(\mathbf{x}' - \mathbf{x}) \right) +$$
$$+ F_{cb}(x')\Big|_{t'=t} \Pi_l(x) \left( g_{qe} \partial_p \delta(\mathbf{x} - \mathbf{x}') - g_{pe} \partial_q \delta(\mathbf{x} - \mathbf{x}') \right). \tag{3.6}$$

Using (3.6): Eq. (3.5) will reduce to

$$\{L_i, L_h\} = \frac{1}{2}(\varepsilon_{ijk}\varepsilon_{leh} - \varepsilon_{iek}\varepsilon_{hjl})\varepsilon_{klm}\varepsilon_{mpq}\int d^3x\ x_j\Pi_e F_{pq} +$$

$$+\frac{1}{2}(\varepsilon_{ijk}\varepsilon_{hgf} - \varepsilon_{ijf}\varepsilon_{hgk})\varepsilon_{klm}\varepsilon_{dlb}\varepsilon_{fed}\varepsilon_{mpq}\int d^3x\ x_j x_g F_{pq}\partial_b\Pi_e =$$

$$= -\frac{1}{2}(\varepsilon_{ihj}\varepsilon_{epq} + \varepsilon_{ihe}\varepsilon_{jpq})\int d^3x\ x_j\Pi_e F_{pq} +$$

$$+\frac{1}{2}(\varepsilon_{ijk}\varepsilon_{hgl} - \varepsilon_{ijl}\varepsilon_{hgk})\varepsilon_{klm}\varepsilon_{mpq}\int d^3x\ x_j x_g F_{pq}\partial_e\Pi_e$$

$$-\frac{1}{2}(\varepsilon_{ijk}\varepsilon_{hgb} - \varepsilon_{ijb}\varepsilon_{hgk})\varepsilon_{kem}\varepsilon_{mpq}\int d^3x\ x_j x_g F_{pq}\partial_b\Pi_e\ .$$

(3.7)

Upon integrating the second and third terms on the right hand side of Eq. (3.7) by parts and simplifying, it will reduce to

$$\{L_i, L_h\} = -\varepsilon_{ihk}L_k + \frac{1}{2}\varepsilon_{ijk}\varepsilon_{hlg}\varepsilon_{klm}\varepsilon_{mpq}\int d^3x\ x_j x_g \Pi_e\partial_e F_{pq} +$$

$$+\frac{1}{2}(\varepsilon_{ijk}\varepsilon_{hgb} - \varepsilon_{ijb}\varepsilon_{hgk})\varepsilon_{kem}\varepsilon_{mpq}\int d^3x\ x_j x_g \Pi_e\partial_b F_{pq} =$$

$$= -\varepsilon_{ihk}L_k + \frac{1}{2}\varepsilon_{ijk}\varepsilon_{hlg}\varepsilon_{klm}\varepsilon_{mpq}\int d^3x\ x_j x_g \Pi_e\partial_e F_{pq} +$$

$$+\frac{1}{2}\varepsilon_{ijr}\varepsilon_{hgs}\varepsilon_{rst}\varepsilon_{tkb}\varepsilon_{kem}\varepsilon_{mpq}\int d^3x\ x_j x_g \Pi_e\partial_b F_{pq} =$$

$$= -\varepsilon_{ihk}L_k - \frac{1}{2}\varepsilon_{ihj}\varepsilon_{mpq}\int d^3x\ x_j x_e \Pi_e\partial_m F_{pq}$$

(3.8)

Eq. (3.8) will reduce, on the constraint surface and on account of $\zeta_4$, to

$$\{L_i, L_h\} \approx -\varepsilon_{ihk}L_k\ ,$$

(3.9)

where Eq. (3.9) is true since the second term in the last equality of Eq. (3.8) vanishes weakly on the constraint surface. This is true because, $\varepsilon_{mpq}\partial_m F_{pq}$, vanishes on account of $\zeta_4$, as we can easily see using Eq. (2.5):

$$\varepsilon_{mpq}\partial_m F_{pq} = \frac{\varepsilon_{mpq}}{a^3 e}\partial_m\left[\phi.(\partial_p\phi\times\partial_q\phi) + \partial_p A_q - \partial_q A_p\right] = \frac{\varepsilon_{mpq}}{a^3 e}\partial_m\phi.(\partial_p\phi\times\partial_q\phi) \approx 0,$$

(3.10)

where we used in the last equality the equation, $\phi.\partial_\mu\phi \approx 0$, which results from the definition of $\zeta_4$. (Where, $\zeta_4 \equiv \phi.\phi - a^2 \approx 0$, see Eq. (2.8).)

The second term on the right hand side of Eq. (3.4) vanishes on the constraint surface. To see that, we start by evaluating the equal-time Poisson brackets of $\zeta_i$'s, and $L_i$'s using Eqs. (2.5,6,8), (3.3):

$$\{L_i(t),\zeta_1(x')\}\big|_{t'=t} = -\frac{1}{2}\varepsilon_{ijk}\varepsilon_{klm}\varepsilon_{mpq}\int d^3x\, x_j \{F_{0l}(x)F_{pq}(x),\zeta_1(x')\big|_{t'=t}\} =$$

$$= -\frac{\varepsilon_{ijk}\varepsilon_{klm}\varepsilon_{mpq}}{2ae}\int d^3x\, x_j F_{0l}(x)\bigg[\partial_{q'}\phi_3(x')\big|_{t'=t}\partial_p\delta(\mathbf{x}-\mathbf{x}') - \partial_{p'}\phi_3(x')\big|_{t'=t}\partial_q\delta(\mathbf{x}-\mathbf{x}') +$$

$$+ \frac{\varepsilon_{3rs}\varepsilon_{ruv}}{a^2}\phi_s(x')\big|_{t'=t}\phi_u(x)\big(\partial_p\phi_v(x)\partial_q\delta(\mathbf{x}-\mathbf{x}') - \partial_q\phi_v(x)\partial_p\delta(\mathbf{x}-\mathbf{x}')\big)\bigg] \approx 0,$$

(3.11)

where (3.11) vanishes on the constraint surface on account of $\zeta_4$.

Similarly,

$$\{L_i(t),\zeta_2(x')\}\big|_{t'=t} = -\frac{1}{2}\varepsilon_{ijk}\varepsilon_{klm}\varepsilon_{mpq}\int d^3x\, x_j \{F_{0l}(x)F_{pq}(x),\zeta_2(x')\big|_{t'=t}\} =$$

$$= -\frac{\varepsilon_{ijk}\varepsilon_{klm}\varepsilon_{mpq}}{2ae}\int d^3x\, x_j F_{0l}(x)\bigg[\partial_{q'}\phi_1(x')\big|_{t'=t}\partial_p\delta(\mathbf{x}-\mathbf{x}') - \partial_{p'}\phi_1(x')\big|_{t'=t}\partial_q\delta(\mathbf{x}-\mathbf{x}') +$$

$$+ \frac{\varepsilon_{1rs}\varepsilon_{ruv}}{a^2}\phi_s(x')\big|_{t'=t}\phi_u(x)\big(\partial_p\phi_v(x)\partial_q\delta(\mathbf{x}-\mathbf{x}') - \partial_q\phi_v(x)\partial_p\delta(\mathbf{x}-\mathbf{x}')\big)\bigg] \approx 0,$$

(3.12)

which also vanishes on the constraint surface on account of $\zeta_4$.

We also, easily, get

$$\{L_i(t),\zeta_3(x')\}\big|_{t'=t} = -\frac{3}{2a^5e}\varepsilon_{ijk}\varepsilon_{mnp}x'_j F_{0l}(x')\phi_m(x')\partial_k\phi_n(x')\partial_l\phi_p(x')\bigg|_{t'=t},$$

$$\{L_i(t),\zeta_4(x')\}\big|_{t'=t} = \{L_i(t),\zeta_5(x')\}\big|_{t'=t} = 0,$$

$$\{L_i(t),\zeta_6(x')\}\big|_{t'=t} = \frac{1}{2}\varepsilon_{ijk}\varepsilon_{k3m}\varepsilon_{mpq}x'_j F_{pq}(x')\phi_3(x')\bigg|_{t'=t},$$

$$\{L_i(t),\zeta_7(x')\}\big|_{t'=t} = \frac{1}{2}\varepsilon_{ijk}\varepsilon_{k3m}\varepsilon_{mpq}x'_j F_{pq}(x')\phi_1(x')\bigg|_{t'=t},$$

$$\{L_i(t),\zeta_8(x')\}\big|_{t'=t} = -\frac{1}{2}\varepsilon_{ijk}\varepsilon_{k3m}\varepsilon_{mpq}x'_j F_{pq}(x')\bigg|_{t'=t}.$$

(3.13)

We see easily, using Eqs. (3.11,12) (which vanish on the constraint surface on account of $\zeta_4$), Eq. (3.13) and the values of $C^{-1}_{\alpha\alpha'}(\mathbf{x},\mathbf{x}';t)$ given in ref.[9], that the second term on the right hand side of Eq. (3.4) vanishes on the constraint surface in a trivial way, since the only non-vanishing elements of $C^{-1}_{\alpha\alpha'}$ are: $C^{-1}_{16}, C^{-1}_{17}, C^{-1}_{18}, C^{-1}_{26}, C^{-1}_{27}, C^{-1}_{28}, C^{-1}_{34}, C^{-1}_{56}, C^{-1}_{57}, C^{-1}_{58}$ and their transposes.

So from the above result and Eq. (3.9) we get,

$$\{L_i, L_h\}_{D(\zeta)} = -\varepsilon_{ihk} L_k, \tag{3.14}$$

*which verifies the first of the Lorentz algebra.*

**b.** Next, to verify the second of the Lorentz algebra by evaluating the Dirac bracket of $K_i$'s:

$$\{K_i(t), K_h(t)\}_{D(\zeta)} \equiv \{K_i(t), K_h(t)\} - \int\int \{K_i(t), \zeta_\alpha(\mathbf{x},t)\} d^3x\, C^{-1}_{\alpha\alpha'}(\mathbf{x},\mathbf{x}';t)\, d^3x' \{\zeta_{\alpha'}(\mathbf{x}',t), K_h(t)\}. \tag{3.15}$$

Where, using Eqs. (2.5,6) and (3.3), and without using any constraints, we get

$$\{K_i(t), K_h(t)\} = \frac{1}{2}\varepsilon_{klm}\varepsilon_{mpq}\int d^3x\, (g_{kn}g_{lh}x_i - g_{kn}g_{li}x_h) F_{0n} F_{pq} =$$

$$= -\frac{1}{2}\varepsilon_{ihj}\varepsilon_{jlk}\varepsilon_{knm}\varepsilon_{mpq}\int d^3x\, x_l F_{0n} F_{pq} = \varepsilon_{ihj} L_j. \tag{3.16}$$

The second term on the right hand side of Eq. (3.15) vanishes on the constraint surface in a trivial way since the only non-vanishing elements of $C^{-1}_{\alpha\alpha'}$ are:

$C^{-1}_{16}, C^{-1}_{17}, C^{-1}_{18}, C^{-1}_{26}, C^{-1}_{27}, C^{-1}_{28}, C^{-1}_{34}, C^{-1}_{56}, C^{-1}_{57}, C^{-1}_{58}$ and their transposes, and since

$$\{K_i(t), \zeta_\alpha(x')\}\Big|_{t'=t} = 0, \quad \text{for } \alpha = 1,2,4,5, \tag{3.17}$$

on the constraint surface on account of $\zeta_4$ alone.

( For the sake of completeness we find:

$$\{K_i(t), \zeta_3(x')\}\Big|_{t'=t} = \frac{3}{4}\varepsilon_{klm} x'_i F_{pq}(x') \phi_k(x') \partial_{p'}\phi_l(x') \partial_{q'}\phi_m(x')\Big|_{t'=t},$$

$$\{K_i(t), \zeta_6(x')\}\Big|_{t'=t} = -x'_i F_{03}(x') \phi_3(x')\Big|_{t'=t},$$

$$\{K_i(t), \zeta_7(x')\}\Big|_{t'=t} = -x'_i F_{03}(x') \phi_1(x')\Big|_{t'=t},$$

$$\{K_i(t), \zeta_8(x')\}\Big|_{t'=t} = x'_i F_{03}(x')\Big|_{t'=t}.$$

$$\tag{3.18}$$

Using Eqs. (3.16) and (3.17), we get:

$$\{K_i, K_h\}_{D(\zeta)} = \varepsilon_{ihk} L_k, \tag{3.19}$$

*which verifies the second of the Lorentz algebra.*

**c.** To verify the next Lorentz algebra by evaluating the equal-time Dirac Bracket of $K_i$'s and $L_h$'s:

$$\{K_i(t), L_h(t)\}_{D(\zeta)} \equiv \{K_i(t), L_h(t)\} - \int\int \{K_i(t), \zeta_\alpha(\mathbf{x},t)\} d^3x\, C^{-1}_{\alpha\alpha'}(\mathbf{x},\mathbf{x}';t)\, d^3x'\, \{\zeta_{\alpha'}(\mathbf{x}',t), L_h(t)\}. \tag{3.20}$$

Using Eqs. (3.3,6) and (2.5,6), The first term on the right hand side of Eq. (3.20) will be

$$\{K_i(t), L_h(t)\} = -\frac{\varepsilon_{hjk}\varepsilon_{klm}\varepsilon_{mpq}}{4}\int d^3x \int d^3x'\, x_i x'_j \left(4g_{nq}F_{0n}(x)F_{0l}(x')\partial_{p'}\delta(\mathbf{x}-\mathbf{x}') + \right.$$

$$\left. + \varepsilon_{rsu}\varepsilon_{rvw}g_{lu}F_{vw}(x)F_{pq}(x')\partial_{s'}\delta(\mathbf{x}-\mathbf{x}')\right) =$$

$$= -\frac{\varepsilon_{hjk}\varepsilon_{klm}\varepsilon_{mpq}}{4}\int d^3x\, x_i \left[4F_{0q}\partial_p(x_j F_{0l}) + \varepsilon_{rsl}\varepsilon_{rvw}F_{vw}\partial_s(x_j F_{pq})\right] =$$

$$= \varepsilon_{hjk}\int d^3x\, x_i x_j F_{0k}\partial_l F_{0l} - \varepsilon_{hjk}\int d^3x\, x_i x_j F_{0l}\partial_k F_{0l} +$$

$$+ \frac{\varepsilon_{hjk}\varepsilon_{klm}}{4}\int d^3x\, x_i x_j F_{lm}\left(\varepsilon_{npq}\partial_n F_{pq}\right) - \frac{\varepsilon_{hjn}\varepsilon_{klm}\varepsilon_{kpq}}{4}\int d^3x\, x_i x_j F_{lm}\partial_n F_{pq},$$

(3.21)

where the first term in the last equality on the right hand side of Eq. (3.21) vanishes weakly on the constraint surface on account of $\zeta_5$ and Eq. (2.6), while the third term on the right hand side vanishes on the constraint surface on account of $\zeta_4$ as was explicitly shown in Eq. (3.10). So, integrating the second and the forth terms on the right hand side by parts and simplifying, Eq. (3.21) will reduce to:

$$\{K_i(t), L_h(t)\} \approx -\varepsilon_{ihj}K_j(t), \tag{3.22}$$

satisfied "weakly" on the constraint surface on account of $\zeta_4$ and $\zeta_5$.

The second term on the right hand side of Eq. (3.20) vanishes on the constraint surface in a trivial way by using Eqs. (3.11,12,13,17) and since the only non-vanishing elements of $C^{-1}_{\alpha\alpha'}$ are: $C^{-1}_{16}, C^{-1}_{17}, C^{-1}_{18}, C^{-1}_{26}, C^{-1}_{27}, C^{-1}_{28}, C^{-1}_{34}, C^{-1}_{56}, C^{-1}_{57}, C^{-1}_{58}$ and their transposes.

So we get:

$$\{K_i(t), L_h(t)\}_{D(\zeta)} = -\varepsilon_{ihj}K_j(t), \tag{3.23}$$

which verifies the last of the homogenous Lorentz algebra.

**d.** Next, we verify the Lorentz algebra involving $P^\mu$. In the monopole's field outside its core (i.e. in the Higgs vacuum region) and by using Eqs. (2.2,4,5,6), Eq. (3.2) and Eq. (13a) in ref.[9] we have

$$P_i = \int d^3x \left(\mathbf{E}_a \times \mathbf{B}_a\right)_i = -\frac{1}{2}\varepsilon_{ilm}\varepsilon_{mpq}\int d^3x\, F_{0l}F_{pq}$$

$$P^0 = H = \frac{1}{2}\int d^3x \left(F_{0j}F_{0j} + \frac{1}{2}F_{pq}F_{pq}\right).$$

(3.24)

Analogously to Eqs. (3.11,12,13), we find on the constraint surface

$$\{P_i(t),\zeta_\alpha(x')\}\big|_{t'=t} = 0, \quad \text{for } \alpha = 1,2,4,5,$$

(3.25)

and

$$\{P_i(t),\zeta_3(x')\}\big|_{t'=t} = -\frac{3}{4a^5 e}\varepsilon_{ijk}\varepsilon_{klm}\varepsilon_{uvw}F_{0j}(x')\phi_u(x')\partial_l\phi_v(x')\partial_m\phi_w(x')\big|_{t'=t}$$

$$\{P_i(t),\zeta_6(x')\}\big|_{t'=t} = \frac{1}{2}\varepsilon_{i3m}\varepsilon_{mpq}F_{pq}(x')\phi_3(x')\big|_{t'=t}$$

$$\{P_i(t),\zeta_7(x')\}\big|_{t'=t} = \frac{1}{2}\varepsilon_{i3m}\varepsilon_{mpq}F_{pq}(x')\phi_1(x')\big|_{t'=t}$$

$$\{P_i(t),\zeta_8(x')\}\big|_{t'=t} = -\frac{1}{2}\varepsilon_{i3m}\varepsilon_{mpq}F_{pq}(x')\big|_{t'=t}.$$

(3.26)

Eqs. (3.11,12,13,17,25) and the fact that the only non-vanishing elements of $C^{-1}_{\alpha\alpha'}$ are $C^{-1}_{16}, C^{-1}_{17}, C^{-1}_{18}, C^{-1}_{26}, C^{-1}_{27}, C^{-1}_{28}, C^{-1}_{34}, C^{-1}_{56}, C^{-1}_{57}, C^{-1}_{58}$ and their transposes imply that *the Dirac brackets of* $P_i$ *with* $P_j$'s, $L_j$'s, *and* $K_j$'s *are equal to the corresponding Poisson brackets evaluated on the constraint surface with constraints,* $\zeta_\alpha$'s, *taken as strong equations.*

So we get using Eqs. (3.24, 6) and integration by parts:

$$\{P_i(t),P_j(t)\}_{D(\zeta)} = \{P_i(t),P_j(t)\}\big|_{\zeta_\alpha\text{'s}=0}$$

$$= \left(\frac{1}{2}\varepsilon_{ijk}\varepsilon_{klm}\int d^3x\, F_{lm}\partial_n\Pi_n - \frac{1}{2}\varepsilon_{ijk}\varepsilon_{mpq}\int d^3x\, \Pi_k\partial_m F_{pq}\right)\bigg|_{\zeta_\alpha\text{'s}=0} = 0,$$

(3.27)

where in the last equality the first term vanishes on account of $\zeta_5$ and the second term vanishes on account of $\zeta_4$ or Eq. (3.10).

Similarly, we also have

$$\{P_i(t),L_j(t)\}_{D(\zeta)} = \{P_i(t),L_j(t)\}\big|_{\zeta_\alpha\text{'s}=0},$$

(3.28)

where using Eqs. (2.6) and (3.3, 6, 24)

$$\{P_i(t),L_j(t)\} = \frac{1}{4}\varepsilon_{ikl}\varepsilon_{lpq}\varepsilon_{jgf}\varepsilon_{fed}\varepsilon_{dcb}\int\int d^3x\, d^3x'\, x'_g \{F_{0k}(x)F_{pq}(x),F_{0e}(x')F_{cb}(x')\}\big|_{t'=t} =$$

$$= \frac{1}{2}\varepsilon_{ilm}\varepsilon_{mpq}\varepsilon_{jkl}\int d^3x\, F_{0k}F_{pq} - \frac{1}{2}\varepsilon_{ilm}\varepsilon_{mpq}\varepsilon_{jgl}\int d^3x\, x_g\, F_{pq}\partial_k F_{0k} +$$
$$+ \frac{1}{2}\varepsilon_{ikl}\varepsilon_{mpq}\varepsilon_{jgl}\int d^3x\, x_g\, F_{pq}\partial_m F_{0k}\,,$$

where the second term on the right hand side vanishes on the constraint surface on account of $\zeta_5$. So, upon integrating the third term on the right hand side by parts and then using Eq. (3.10), which results from $\zeta_4$, we get on the constraint surface

$$\{P_i(t), L_j(t)\} \approx \frac{1}{2}\varepsilon_{ijk}\varepsilon_{klm}\varepsilon_{mpq}\int d^3x\, F_{0l}F_{pq} = -\varepsilon_{ijk}P_k(t)\,,$$

which implies when substituting in Eq. (3.28)

$$\{P_i(t), L_j(t)\}_{D(\zeta)} = -\varepsilon_{ijk}P_k(t)\,. \qquad (3.29)$$

We also have

$$\{P_i(t), K_j(t)\}_{D(\zeta)} = \{P_i(t), K_j(t)\}\big|_{\zeta_\alpha\text{'s}=0}\,, \qquad (3.30)$$

where using Eqs. (2.5, 6) and (3.3, 24) we have

$$\{P_i(t), K_j(t)\} = \varepsilon_{ilm}\varepsilon_{mpq}\int\int d^3x\, d^3x'\, x'_j\left[F_{0k}(x')\big|_{t'=t} F_{0l}(x)\, g_{qk}\partial_p \delta(\mathbf{x}-\mathbf{x}') +\right.$$
$$\left. + \frac{\varepsilon_{ruv}\varepsilon_{rsw}}{4} F_{uv}(x')\big|_{t'=t} F_{pq}(x)g_{sl}\partial_{w'}\delta(\mathbf{x'}-\mathbf{x})\right] =$$
$$= \int d^3x\, x_j\, F_{0k}\partial_i F_{0k} - \int d^3x\, x_j\, F_{0i}\partial_k F_{0k} - \frac{\varepsilon_{ilm}\varepsilon_{mpq}\varepsilon_{kuv}}{4}\int d^3x\, F_{pq}\left(\varepsilon_{klj} - \varepsilon_{klw}x_j\partial_w\right)F_{uv}\,,$$

**(3.31)**

where the second term on the right hand side of the last equality will vanish on the constraint surface on account of $\zeta_5$.

Integrating the first term on the right hand side of Eq. (3.31) by parts and simplifying the third term, and then integrating one of its resulting terms further by parts and simplifying further:

$$\{P_i(t), K_j(t)\} \approx \frac{1}{2}\delta_{ij}\int d^3x\, F_{0k}F_{0k} + \frac{\delta_{ij}\varepsilon_{klm}\varepsilon_{mpq}}{8}\int d^3x\, F_{kl}F_{pq} - \frac{\varepsilon_{ikl}}{4}\int d^3x\, x_j F_{kl}\left(\varepsilon_{mpq}\partial_m F_{pq}\right),$$

and the third term on the right hand side will vanish on account of Eq. (3.10), or equivalently $\zeta_4$. So we get using Eq. (3.24)

$$\{P_i(t), K_j(t)\} \approx \delta_{ij}H(t)\,,$$

which if substituted in Eq. (3.30) implies

$$\{P_i(t), K_j(t)\}_{D(\zeta)} = \delta_{ij} H(t). \tag{3.32}$$

**e.** Finally, the Lorentz algebra involving H:

Analogously to Eqs. (3.11,12,13), we find on the constraint surface

$$\{H(t), \zeta_\alpha(x')\}\big|_{t'=t} = 0, \quad \text{for } \alpha = 1, 2, 4, 5, \tag{3.33}$$

and

$$\{H(t), \zeta_3(x')\}\big|_{t'=t} = \frac{3}{a^5 e} \varepsilon_{klm} F_{pq}(x') \phi_k(x') \partial_p \phi_l(x') \partial_q \phi_m(x')\big|_{t'=t}$$

$$\{H(t), \zeta_6(x')\}\big|_{t'=t} = -F_{03}(x') \phi_3(x')\big|_{t'=t}$$

$$\{H(t), \zeta_7(x')\}\big|_{t'=t} = -F_{03}(x') \phi_1(x')\big|_{t'=t}$$

$$\{H(t), \zeta_8(x')\}\big|_{t'=t} = F_{03}(x')\big|_{t'=t}.$$

$$\tag{3.34}$$

Eqs. (3.11,12,13,17,25,33) and the fact that the only non-vanishing elements of $C^{-1}_{\alpha\alpha'}$ are $C^{-1}_{16}, C^{-1}_{17}, C^{-1}_{18}, C^{-1}_{26}, C^{-1}_{27}, C^{-1}_{28}, C^{-1}_{34}, C^{-1}_{56}, C^{-1}_{57}, C^{-1}_{58}$ and their transposes imply that *the Dirac brackets of H with $P_j$'s, $L_j$'s, and $K_j$'s are equal to the corresponding Poisson brackets evaluated on the constraint surface with constraints, $\zeta_\alpha$'s, taken as strong equations.*

So we have

$$\{L_i(t), H(t)\}_{D(\zeta)} = \{L_i(t), H(t)\}\big|_{\zeta_\alpha \text{'s}=0}, \tag{3.35}$$

where using Eqs. (2.5,6) and (3.3,24)

$$\{L_i(t), H(t)\} = -\frac{\varepsilon_{ijk}\varepsilon_{klm}\varepsilon_{mpq}}{4} \left( \int\int d^3x\, d^3x'\, x_j \{F_{0l}(x)F_{pq}(x), F_{0n}(x')F_{0n}(x')\}\big|_{t'=t} + \right.$$

$$\left. + \frac{\varepsilon_{bcd}\varepsilon_{bfg}}{4} \int\int d^3x\, d^3x'\, x_j \{F_{0l}(x)F_{pq}(x), F_{cd}(x')F_{fg}(x')\}\big|_{t'=t} \right) =$$

$$= \varepsilon_{ijk}\varepsilon_{klm}\varepsilon_{mpq} \int\int d^3x\, d^3x'\, x_j \left( F_{0n}(x')\big|_{t'=t} F_{0l}(x) g_{qn} \partial_p \delta(\mathbf{x}-\mathbf{x}') + \right.$$

$$\left. + \frac{\varepsilon_{bcd}\varepsilon_{bfg}}{4} F_{fg}(x')\big|_{t'=t} F_{pq}(x) g_{cl} \partial_{d'} \delta(\mathbf{x}'-\mathbf{x}) \right),$$

$$\tag{3.36}$$

where upon integrations by parts in suitable places, using the properties of the Levi Civita tensor, and simplifying: Eq. (3.36) will reduce to

$$\{L_i(t), H(t)\} = -\varepsilon_{ijk} \int d^3x\, x_j F_{0k}(\partial_l F_{0l}) - \frac{\varepsilon_{ijk}\varepsilon_{klm}}{4} \int d^3x\, x_j F_{lm}(\varepsilon_{npq}\partial_n F_{pq}) \approx 0, \tag{3.37}$$

where the first term on the right hand side vanishes on the constraint surface on account of $\zeta_5$, and the second term vanishes on account of $\zeta_4$ or Eq. (3.10). So, Eq. (3.35) will now give

$$\{L_i(t), H(t)\}_{D(\zeta)} = \{L_i(t), H(t)\}\big|_{\zeta_\alpha's=0} = 0. \tag{3.38}$$

We also have

$$\{K_i(t), H(t)\}_{D(\zeta)} = \{K_i(t), H(t)\}\big|_{\zeta_\alpha's=0}, \tag{3.39}$$

where using Eqs. (2.5,6) and (3.3,24)

$$\{K_i(t), H(t)\} = \frac{\varepsilon_{klm}\varepsilon_{mpq}}{16} \int\int d^3x\, d^3x'\, x_i \Big(\{F_{0j}(x)F_{0j}(x), F_{kl}(x')F_{pq}(x')\}\big|_{t'=t} +$$

$$+ \{F_{kl}(x)F_{pq}(x), F_{0j}(x')F_{0j}(x')\}\big|_{t'=t}\Big) =$$

$$= \frac{\varepsilon_{klm}\varepsilon_{mpq}}{2} \int\int d^3x\, d^3x'\, x_i \Big(F_{pq}(x')\big|_{t'=t} F_{0j}(x)\, g_{jl}\partial_{k'}\delta(\mathbf{x}-\mathbf{x}') +$$

$$+ F_{0j}(x')\big|_{t'=t} F_{kl}(x) g_{jp}\partial_q \delta(\mathbf{x}'-\mathbf{x})\Big) =$$

$$= \frac{1}{2}\varepsilon_{ikl}\varepsilon_{lpq} \int d^3x\, F_{0k} F_{pq} = -P_i. \tag{3.40}$$

So Eq. (3.39,40) now give

$$\{K_i(t), H(t)\}_{D(\zeta)} = \{K_i(t), H(t)\}\big|_{\zeta_\alpha's=0} = -P_i. \tag{3.41}$$

Finally we also have

$$\{P_i(t), H(t)\}_{D(\zeta)} = \{P_i(t), H(t)\}\big|_{\zeta_\alpha's=0}, \tag{3.42}$$

where using Eqs. (2.5,6) and (3.24)

$$\{P_i(t), H(t)\} = \frac{\varepsilon_{iml}\varepsilon_{mpq}}{4} \int\int d^3x\, d^3x' \Big\{F_{0l}(x)F_{pq}(x), F_{0j}(x')F_{0j}(x') + \frac{\varepsilon_{krs}\varepsilon_{kuv}}{4} F_{rs}(x')F_{uv}(x')\Big\}\Big|_{t'=t} =$$

$$= \varepsilon_{ilm}\varepsilon_{mpq} \int\int d^3x\, d^3x' \Big[F_{0j}(x')\big|_{t'=t} F_{0l}(x) g_{jq}\partial_p \delta(\mathbf{x}-\mathbf{x}') +$$

$$+ \frac{\varepsilon_{krs}\varepsilon_{kuv}}{4} F_{uv}(x')\big|_{t'=t} F_{pq}(x) g_{lr}\partial_{s'}\delta(\mathbf{x}'-\mathbf{x})\Big],$$

where it reduces, upon integrations by parts in suitable places, using the properties of the Levi Civita tensor, dropping the surface terms at infinity and simplifying, to:

$$\{P_i(t), H(t)\} = -\int d^3x\, F_{0i}\left(\partial_j F_{0j}\right) - \frac{\varepsilon_{ijk}}{4} \int d^3x\, F_{jk}\left(\varepsilon_{mpq}\partial_m F_{pq}\right) \approx 0, \tag{3.43}$$

where the first term on the right hand side vanishes on the constraint surface on account of $\zeta_5$, and the second term vanishes on account of $\zeta_4$ or Eq. (3.10). So, Eq. (3.42) will now give

$$\{P_i(t), H(t)\}_{D(\zeta)} = \{P_i(t), H(t)\}\big|_{\zeta_\alpha \text{'s}=0} = 0. \quad (3.44)$$

*Eqs. (3.14,19,23,27,29,32,38,41,44) are strong equations, since inside the Dirac brackets the constraints equations are taken to be strong. Hence, if the Lorentz algebra is valid at the first level of Dirac brackets then it will also be valid at all higher levels.*

## 4. Conclusion

While ref.[4] showed that the Lorentz invariance of non-Abelian monopoles to be broken at the "classical" level, Eqs. (3.14,19,23,27,29,32,38,41,44) here show explicitly that en route to "quantization", we were able to restore the Lorentz invariance of the 't Hooft-Polyakov monopole field. Here we used recent results from the Dirac quantization of the 't Hooft-Polyakov monopole field (i.e. in the Higgs vacuum), given by ref.[9], to show that the Lorentz algebra is valid in this region upon quantization. In particular, we used the constraints $\zeta_4$ and $\zeta_5$ repeatedly in evaluating the Dirac brackets of the Lorentz algebra here. While $\zeta_4$ is just the Higgs vacuum condition, it seemed that $\zeta_5$ was most essential in proving the Lorentz invariance in this region.

## Acknowledgment

I thank the Ilfat & Bah.-Foundation (ed'Oreen, Btouratij) for their continuous support. I thank Prof. Sudarshan for offering the problem, reference [4], and guidance.

xvii